# Giant magneto-birefringence effect and tuneable colouration of 2D crystals' suspensions

Baofu Ding[1], Wenjun Kuang[2], Yikun Pan[1], I. V. Grigorieva[2], A. K. Geim[1,2], Bilu Liu[1], Hui-Ming Cheng[1,3,4]

[1]Environmental Science and New Energy Technology Center, Tsinghua-Berkeley Shenzhen Institute, Tsinghua University, 1001 Xueyuan Road, Shenzhen 518055, China
[2]National Graphene Institute, University of Manchester, Manchester M13 9PL, UK
[3] Shenyang National Laboratory for Materials Science, Institute of Metal Research, Chinese Academy of Sciences, Shenyang 110016, China
[4] Advanced Technology Institute (ATI), University of Surrey, Guildford, Surrey GU2 7XH, UK

**One of the long sought-after goals in manipulation of light through light-matter interactions[1-5] is the realization of magnetic-field-tuneable colouration – so-called magneto-chromatic effect[6] – which holds great promise for optical, biochemical and medical applications due to its contactless and non-invasive nature. This goal can be achieved by magnetic-field controlled birefringence, where colours are produced by the interference between phase-retarded components of transmitted polarised light. Thus far birefringence-tuneable coloration has been demonstrated using electric field[7,8], material chirality[4] and mechanical strain[9] but magnetic field control remained elusive due to either weak magneto-optical response of transparent media[10,11] or low transmittance to visible light of magnetically responsive media, such as ferrofluids[12]. Here we demonstrate magnetically tuneable colouration of aqueous suspensions of two-dimensional cobalt-doped titanium oxide which exhibit an anomalously large magneto-birefringence effect. The colour of the suspensions can be tuned over more than two wavelength cycles in the visible range by moderate magnetic fields below 0.8 T. We show that such giant magneto-chromatic response is due to particularly large phase retardation (>3π) of the polarised light, which in its turn is a combined result of a large Cotton-Mouton coefficient (three orders of magnitude larger than for known liquid crystals), relatively high saturation birefringence ($\Delta n_s \approx$ 2x10⁻⁴) and high transparency of our suspensions to visible light. The work opens a new avenue to achieve tuneable colouration through engineered magnetic birefringence and can readily be extended to other magnetic 2D nanocrystals[13-16]. The demonstrated effect can be used in a variety of magneto-optical applications, including magnetic field sensors, wavelength-tuneable optical filters and see-through printing.**

So far, most attempts to achieve magnetically-controlled coloration have been based on diffraction of light propagating through a periodic structure of suspended nanoparticles (colloidal photonic crystal)[6,17-19], where the characteristic wavelength $\lambda$ is determined by the Bragg condition $\lambda = 2nd\sin\theta$ (here $n$ is the refractive index of water, $d$ the particle spacing). If $d$ can be modified by an external magnetic field, the wavelength of diffracted light can be engineered accordingly. However, magneto-colouration by this method is usually non-uniform due to the difficulty of forming a stable uniform periodic structure of the nanoparticles[6,17-19]. Alternatively, the colour can be manipulated using polarized light as it transmits through a birefringent medium sandwiched between two polarizers. Under the crossed-polarizer configuration, the intensity of transmitted light $I$ is related to the wavelength as $I = \sin^2 \frac{\delta}{2}$, where $\delta = \frac{2\pi \Delta n L}{\lambda}$ is the phase retardation between two polarised light components, $\Delta n$ the birefringence and $L$ the length of the optical path through the medium. With a suitable combination of $\Delta n$ and $L$, the wavelength-dependent intensity can be modulated to achieve different colours. Although this mechanism of spectral tuning has been known for a long time, its applications were limited to measuring the birefringence ($\Delta n$) of different media or the crystal thickness[20,21]. To the best of our knowledge, no attempt has ever been made to use it in order to achieve magneto-chromatic control, likely due to the difficulty of achieving sufficiently large phase retardation $\delta$: Our analysis (see below) shows that, to observe colouration in a birefringent system, $\delta \geq 3\pi$ is needed for λ in the visible range, which is not easily satisfied even in magnetically highly responsive systems, such as ferrofluids, where the achievable $\delta$ is generally below π due to opacity (limited $L$) and small $\Delta n$ due to the weak shape anisotropy of suspended ferromagnetic nanoparticles[11,12,22,23]. Two-dimensional (2D) crystals, on the other hand, offer excellent prospects for magneto-birefringent colouration, thanks to their large shape anisotropy and a favourable combination of optical and magnetic properties[13,15,16,24-26]. In the case of aqueous suspensions of two-dimensional cobalt-doped titanium oxide (2D Co-TiO$_x$), their appreciable magnetic anisotropy (largely due to the known single-ion anisotropy of Co$^{2+}$ ions in crystalline environments[27]) enables tunability by the magnetic field, while the relatively large birefringence ($\Delta n \sim 10^{-4}$ at 0.02% volume concentration) and > 50% transparency to visible light result in a large $\delta \approx 6\pi$ at $\mu_0 H$ = 800 mT, allowing magnetic colouration that covers more than two wavelength cycles in the visible.

Suspensions of 2D Co-TiO$_x$ were prepared by liquid exfoliation of bulk Co-doped TiO$_x$ crystals in water using mechanical agitation. Synthesis and characterisation of the bulk crystals is described in Methods and Extended Data Fig. 1; briefly, bulk Co-TiO$_x$ was synthesized from a mixture of TiO$_2$, CoO, K$_2$CO$_3$ and Li$_2$CO$_3$ using an annealing and ion-exchange method. 2D Co-TiO$_x$ crystals obtained after exfoliation have a lepidocrocite-type structure where Co$^{2+}$ ions substitute some of Ti$^{4+}$ at the octahedral sites as schematically shown in Fig. 1a. The crystals have a large aspect ratio with a typical lateral size of ~1.5 µm and thickness of ~1 nm (Extended Data Fig. 1c-e). Most of the results presented below have been obtained on a 2D Co-TiO$_x$ suspension with a volume concentration of 0.02 vol%, exhibiting a high transmittance over visible λ (Extended Data Fig. 1f); qualitatively similar results were also obtained on more dilute suspensions. Our optical setup is sketched in Fig.1a: a tall quartz cuvette filled with a suspension of 2D crystals was placed between two crossed polarizers in a gradient magnetic field applied in the vertical direction, increasing from bottom to top (see Methods for details of optical measurements). When back-lit with white light, a vertical rainbow of colours appeared in the cuvette, with different colours corresponding to different field strengths (Fig. 1b, right and Supplementary Video 1). To quantify the field-colour correspondence, the same suspension was placed between crossed polarizers in a uniform magnetic field. For low fields $\mu_0 H < 325$ mT, as the magnetic field increased from $\mu_0 H = 0$, the suspension gradually became more transparent, but no colour was observed (Extended Data Fig. 2). As the field increased above 325 mT, a colour appeared and started to evolve first from yellow to orange, then purple to blue, and finally to green at the highest field, $\mu_0 H \approx 800$ mT (Fig. 1c and Supplementary Video 1).

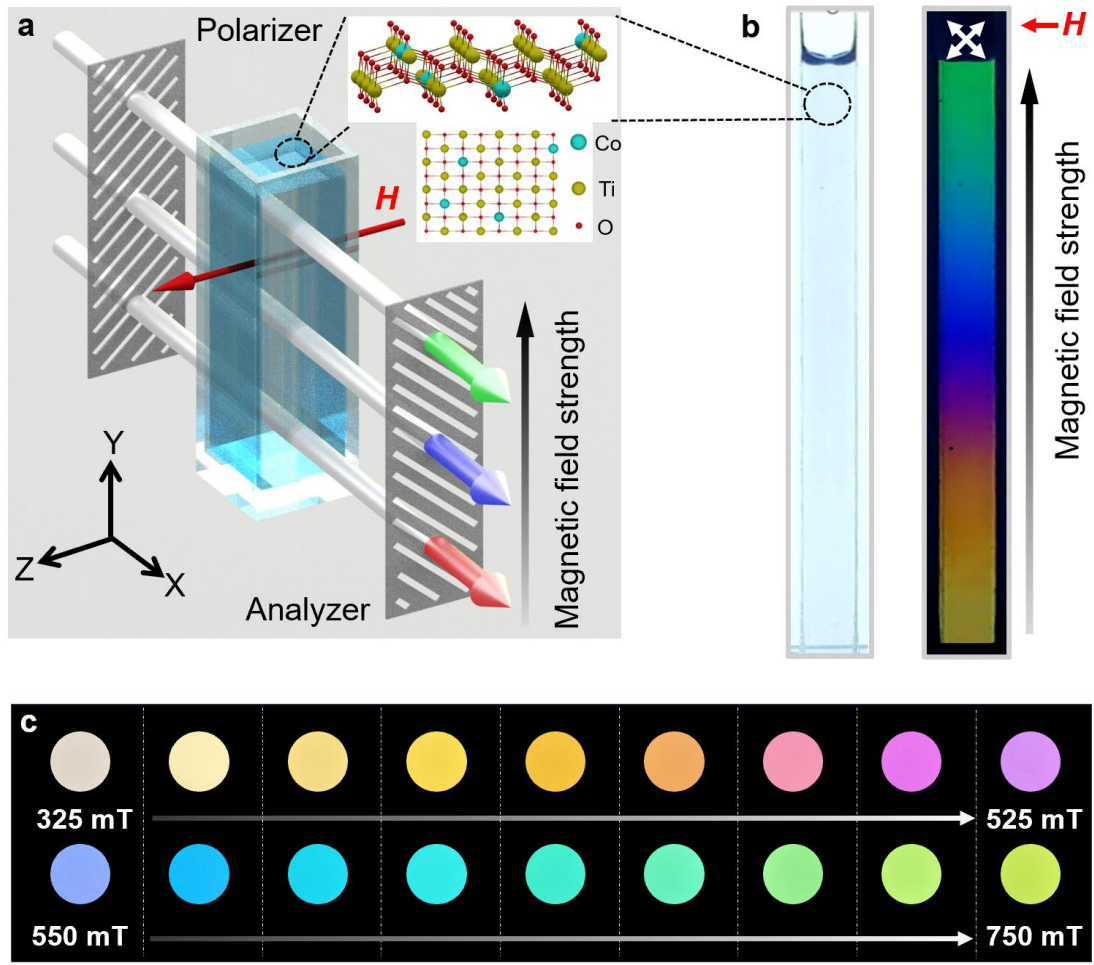

**Fig. 1| Magneto-chromatic effect in a 2D Co-TiO$_x$ aqueous suspension. a**, Schematic setup of our magneto-optical experiments. A 5.5 cm tall cuvette with 1.0 x 0.5 cm rectangular cross-section filled with a suspension of 2D Co-TiO$_x$ crystals is placed between two crossed polarizers, with the magnetic field applied along the z direction, at 45° to the polariser/analyser. The top-right inset shows schematic atomic structure of 2D Co-TiO$_x$. The light path is along the *x* direction and the magnetic field is increasing vertically from bottom to top. **b**, Real images of the cuvette filled with 0.02 vol% suspensions and illuminated with uniform white light. Left: no colours appear if the polariser/analyser are absent; right: a rainbow of colours is clearly seen in the presence of crossed polarizers. **c**, Real images of the colours of the suspension in uniform magnetic field as the field is tuned from $\mu_0 H = 325$ mT to 750 mT with a 25 mT step.

The observed magneto-coloration cannot be the result of Bragg diffraction by a periodic pattern of suspended 2D crystals, because the colour disappeared when the polarizers were removed but the field kept on (Fig. 1b, left). Instead, as we show below, colouration is the result of magnetically induced birefringence of the suspension. In the absence of magnetic field, 2D Co-TiO$_x$ crystals are randomly oriented in the water, exhibiting an isotropic optical response with zero birefringence, $\Delta n = 0$, and the suspension remains opaque. With the field applied, the crystals start to rotate and align parallel to *H* due to their magnetic anisotropy, collectively giving rise to a finite optical anisotropy (birefringence). The mechanism responsible for the magnetically induced birefringence is illustrated in Fig. 2a. Before passing through the suspension, the linearly polarised white source can be represented by three primary colours with wavelengths $\lambda_1$, $\lambda_2$ and $\lambda_3$, of equal amplitudes. As the magnetic field is applied, 2D Co-TiO$_x$ crystals rotate and align parallel to the field, resulting in a finite birefringence of the suspension, such that the two polarised light components (parallel and perpendicular to **H**) experience a wavelength-dependent phase retardation $\delta = 2\pi\Delta n L/\lambda$. When passing through the analyzer, constructive interference between these two components occurs for $\delta = (2N - 1)\pi$, leading to a transmission maximum ($N = 1,2,3$ *etc.* is an integer). Due to the wavelength dependence of $\delta$, if the $\lambda_2$ component transmits constructively, the $\lambda_1$ and $\lambda_3$ components are off resonance and have lower transmittance, resulting in a colour dominated by $\lambda_2$.

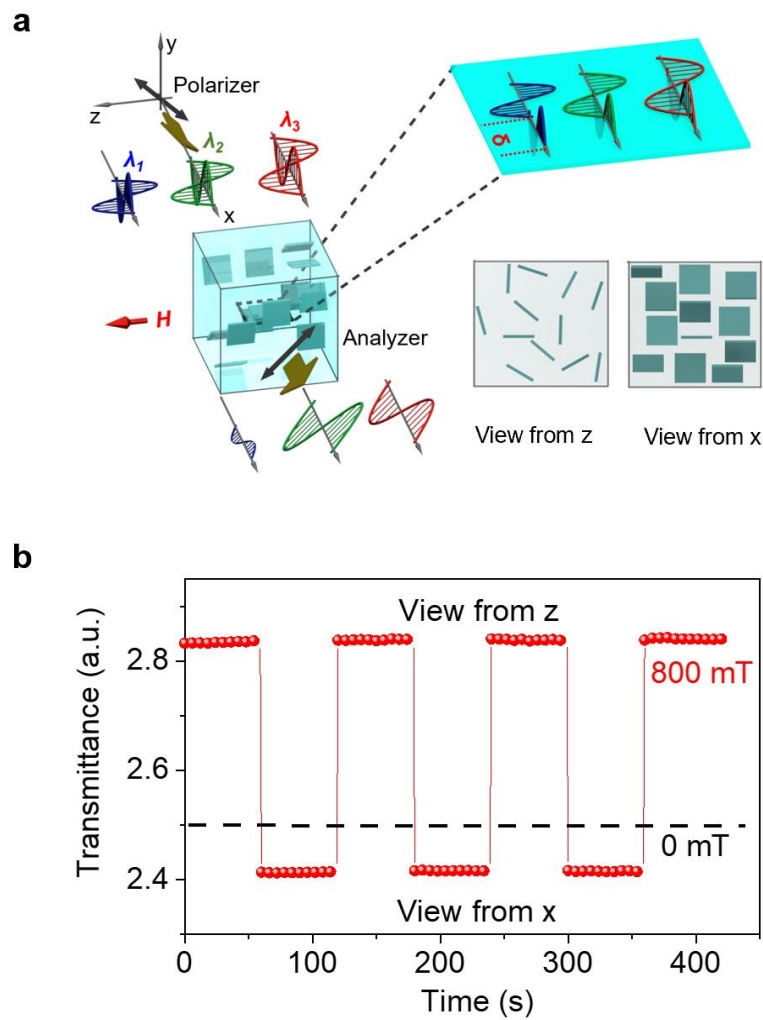

**Fig. 2| Alignment of suspended 2D Co-TiO$_x$ crystals by the magnetic-field. a**, Schematic explanation of the colouration by magnetic-field induced birefringence of the 2D Co-TiO$_x$ suspension (see text). **b,** Transmittance anisotropy of the 2D Co-TiO$_x$ suspension under a uniform magnetic field $\mu_0 H$ = 800 mT for a 650 nm light wave passing along the field direction (view from *z*) and perpendicular to the field (view from *x*). Less scattering (higher transmission) along the field direction indicates parallel alignment of 2D Co-TiO$_x$ crystals by the magnetic field.

Quantitatively, the relation between $\Delta n$ and $H$ can be derived using a microscopic model analogous to the electro-birefringence (Kerr) effect[28]:

$$\Delta n(H) = \frac{\Delta n_s}{2}\left[3L_2\left(\frac{\Delta\chi H^2}{2k_B T}\right) - 1\right] \qquad (1)$$

where $L_2(x)$ is the 2$^{nd}$ order generalised Langevin function, $\Delta\chi = \chi_\parallel - \chi_\perp$ the anisotropy of the magnetic susceptibility, $k_B$ the Boltzmann constant and $T$ the temperature (see Supplementary information for a detailed description of our model). It follows from (1) that the birefringence increases proportionally to $H^2$ until saturation at $\Delta n_s$, corresponding to full alignment of all crystals in the suspension parallel to the magnetic field at a sufficiently high $H$. It follows that the wavelength-dependent condition for constructive interference $\delta = \frac{2\pi\Delta n(H)L}{\lambda_c} = (2N - 1)\pi$ also changes with $H$, and the colour of the transmitted light is determined by $\lambda_c$, the constructive interference wavelength.

We have verified the magnetic alignment/rotation of suspended Co-TiO$_x$ crystals by comparison of the magneto-optical transmittance when viewed along and perpendicular to the applied field (in the direction of the *z*-axis and *x*-axis in Fig. 1a, respectively). As demonstrated in Fig. 2b, the transmittance is significantly higher in the former case, in agreement with a higher in-plane magnetization which aligns the flakes along the field direction, so that they are parallel to the light path and allow more light through. Conversely, when the light path is perpendicular to the magnetic field, it is blocked by the suspended crystals, similar to window blinds. The magnetic anisotropy concluded from the optical experiments is further verified by measurements of in-plane vs out-of-plane magnetisation of our 2D Co-TiO$_x$ crystals, showing a notably higher magnetic susceptibility for **H** parallel to their surfaces (see Extended Data Fig. 3 and Supplementary information).

For a quantitative characterisation of the observed magneto-chromatic effect, we recorded transmitted spectra for different wavelengths of the incident light, at different magnetic fields. The resulting map of the transmitted intensity as a function of *H* and λ is shown in Fig. 3a. As expected from the model outlined above, the transmitted intensity shows oscillations as a function of both *H* and λ, with three transmittance maxima (white stripes in Fig. 3a) and minima (dark-grey stripes) appearing alternately within the field range of 0 to 800 mT. Such a sinusoidal-like behaviour of the transmittance is in excellent agreement with the relation $I \propto \sin^2 \frac{\pi \Delta n(H) L}{\lambda}$, where the intensity maxima correspond to the constructive interference condition given above and intensity minima to the destructive interference with $\delta = \frac{2\pi \Delta n(H) L}{\lambda} = (2N - 2)\pi$. Here $N$ is the $N^{th}$ order of the intensity maximum/minimum. Several individual spectra at different $H$ are shown in Fig. 3b, where the red-shift of the spectra by the magnetic field is clearly seen. For clarity, only 2nd order maxima are shown, corresponding to the coloured dots in Fig. 3a. It is this magnetic field-induced red shift of the spectra that is responsible for the tuneable colouration shown in Fig. 1. A typical dependence of the transmitted intensity on the magnetic field is shown in Fig. 4a: The values of $H$ corresponding to minima and maxima of the light transmission (marked by circles in the figure) correspond to the conditions of destructive/constructive interference, respectively, and were used to analyse the dependence of the phase retardation $\delta$ on $H$ for each wavelength of the incident light (Fig. 4b). Using equation (1) for $\Delta n(H)$, the intensity $I \propto \sin^2 \frac{\pi \Delta n(H) L}{\lambda}$, and the dispersion of $\Delta n(\lambda)$ in the visible range (Fig. 4c), we simulated the $I$ vs. $\lambda, H$ map shown in Fig. 3c. Modelling details can be found in the Supplementary Information. The calculated intensity map is in excellent agreement with the experimental data in Fig. 3a. Furthermore, we derived an intensity map from direct measurements of the phase retardation $\delta(H, \lambda)$ using spectroscopic ellipsometry, as $I = \sin^2 \frac{\delta}{2}$. This yielded an identical transmitted intensity distribution to Fig. 3a (see Extended Data Fig. 4).

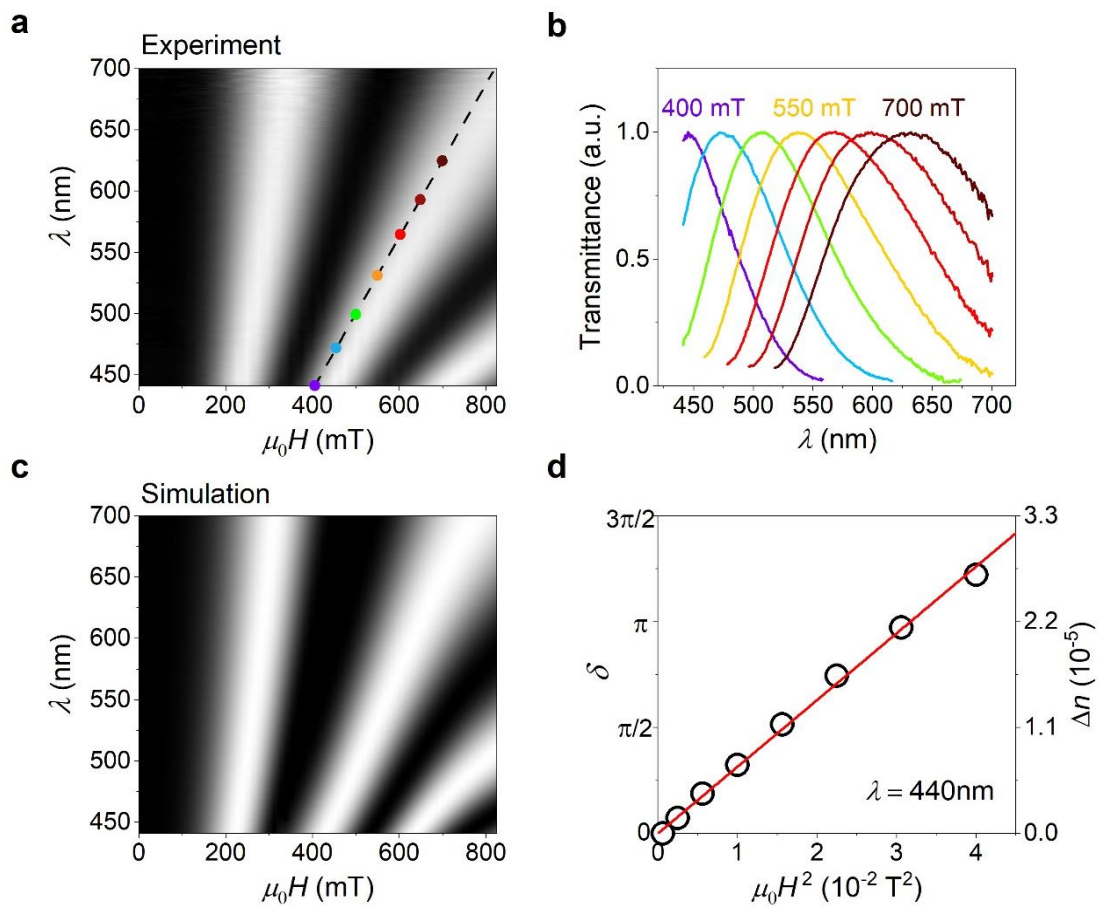

**Fig. 3| Magnetic field-dependent transmission of light through the 2D Co-TiOx suspension at different wavelengths. a**, Experimental map of the normalized transmittance as a function of λ and *H*. Gray scale: black for zero intensity and white for the maximum intensity. **b**, Red shift of the constructive interference wavelength $\lambda_c$ (corresponding to transmittance maxima) with *H*. The wavelength maxima for the colour-coded spectra correspond to the coloured dots in (a). **c**, Simulation of the transmitted light intensity based on our statistical model of magnetic-field induced birefringence (see text). Gray scale: black for zero intensity and white for the maximum intensity. Bright and dark stripes correspond to $\delta(H, \lambda) = (2N - 1)\pi$ and $(2N - 2)\pi$, respectively ($N = 1,2,3$). **d,** Magnetic-field dependence of the phase retardation and birefringence in the low-field range, 0 to 200 mT. Symbols: experimental data; solid red line: linear fit. Here $\Delta n$ is calculated from the measured $\delta(H)$ for λ = 440 nm as $\Delta n(H) = \frac{\lambda}{2\pi L}\delta(H)$.

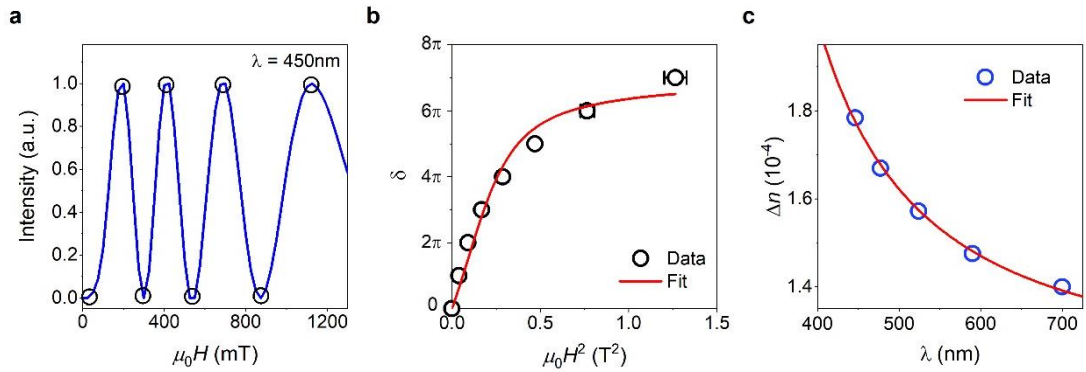

**Fig. 4| Analysis of magnetic field-induced birefringence of 2D Co-TiO$_x$ suspension. a**, Magnetic field dependence of the transmitted light intensity for λ = 450 nm. **b**, Magnetic field dependence of phase retardation for λ = 450 nm. Symbols are data points corresponding to the conditions of either constructive or destructive interference at a particular field $H$; for example δ = π corresponds to the first maximum of intensity in (a), δ = 2π to the second minimum and so on. Solid red line is the fit to the theoretical dependence $\delta(H) = \frac{2\pi \Delta n(H) L}{\lambda}$, where $\Delta n(H)$ is given by (1). **c**, Dispersion of the birefringence. Symbols show $\Delta n$ (λ) calculated using $\Delta n = \lambda \delta / 2\pi L$ where each value of λ corresponds to either minimum or maximum transmission intensity measured at $\mu_0 H$ = 1400 mT. Solid red line is a fit to the wavelength dependence of $\Delta n$ derived in section 3 of the Supplementary information.

Using the data for $\Delta n(H)$ extracted from the transmitted intensity measurements for low magnetic fields, 0 to 200 mT (Fig. 3d), we were able to evaluate the Cotton-Mouton (CM) coefficient for our Co-TiO$_x$ suspensions, $C = \frac{1}{\lambda}\frac{\partial \Delta n}{\partial (H^2)}$. As is clear from the magnetic field dependence of the phase retardation $\delta$ (Extended Data Fig. 4b), at low magnetic fields the first term in (1) is linear with $H^2$ (magnetic susceptibilities are constant at room temperature) and the CM coefficient can be found as $\approx \frac{1}{\lambda}\frac{\Delta n}{H^2}$, yielding $C$ as large as 1400 T$^{-2}$m$^{-1}$ for $\lambda$ = 440 nm, i.e., three orders of magnitude larger than for, e.g., organic liquid crystals (c.f. Extended Data Table). Such a large CM coefficient explains the particularly strong magnetic response of the 2D Co-TiO$_x$ suspensions in our experiments, allowing tunability of the colouration by moderate magnetic fields.

It is instructive to discuss the conditions required to observe magneto-colouration in a birefringent system. Detailed analysis (Supplementary Information) shows that, to achieve magnetic tunability of the colour of transmitted light, one needs to drive the birefringent system to at least the 2$^{nd}$ order of the constructive interference condition, which corresponds to the phase retardation $\delta \geq 3\pi$ (central white stripe in the intensity maps of Fig. 3a,c). This is because for the 1$^{st}$ order spectral maxima (first white stripe on the left of Fig. 3a,c, where 150 mT < $\mu_0 H$ < 325 mT) the spectra of transmitted light at a single $H$ is only weakly dependent with the wavelength (Extended Data Fig. 2b) to allow discrimination of colours, and one can only detect the change of brightness of the transmitted white light, as observed in our experiments at $\mu_0 H$ < 325 mT (Extended Data Fig. 2a). This also explains why in experiments with ferrofluids[12,29], the samples became brighter (more transparent) but no colour could be observed, despite their high magnetic responsivity. Furthermore, it follows from the expression for the phase retardation $\delta = \frac{2\pi \Delta n L}{\lambda}$ that, to achieve $\delta \geq 3\pi$ at visible wavelengths, requires a sufficiently large product $\Delta n L$. This is difficult to achieve in ferrofluids because of their typically small birefringence (an order of magnitude smaller than our 2D Co-TiO$_x$ suspensions with $\Delta n_s \sim 2 \times 10^{-4}$) and low transparency (optical path length $L$ well below 1 cm). We can therefore conclude that, to achieve magneto-colouration, a birefringent suspension must simultaneously meet three conditions: (1) sufficiently large magnetic anisotropy of the suspended crystals, to allow magnetic tunability of the birefringence at easily accessible fields, (2) sufficiently large birefringence $\Delta n$ and (3) high transparency to visible light (sufficiently long $L$). This analysis shows that the effect demonstrated here can be readily extended through either an increased volume fraction of the suspended crystals (Extended Data Fig. 5) or a longer optical path length (Extended Data Fig. 6), and should also be possible to achieve in other 2D magnetic suspension systems, e.g., suspensions of 2D ferromagnetic crystals, provided that they are sufficiently transparent to visible light.

The pronounced magneto-chromatic effect for 0.02 vol% suspension of 2D Co-TiO$_x$ clearly demonstrates the feasibility of achieving magnetically controlled colouration using suspensions of 2D crystals as a birefringent platform. Our additional experiments with more dilute suspensions of Co-TiO$_x$ and different lengths of the optical path $L$ (Extended Data Figs 5 and 6) showed that coloration can still be achieved for suspensions as dilute as 0.01 vol% and the effect is strongly enhanced by increasing $L$, for example, by increasing the thickness of the liquid layer through which the light passes. In terms of potential applications of 2D Co-TiO$_x$ suspensions, we note that the magneto-chromatic effect was found to be highly reproducible, no degradation was observed over time and the suspensions have been stable against restacking or agglomeration for over two years since the start of this work. Furthermore, the possibility to use a longer optical path ensures sufficient tolerance against small variations of the optical length, which is important to achieve uniform colouration. Finally, our preliminary experiment (Extended Data Fig. 7) showed that the time needed to achieve a new colour does not exceed 30 millisecond. Based on the combination of these favourable characteristics, one can envisage that suspensions of magnetically responsive 2D crystals are likely to find applications in a range of devices, including magnetic displays, magnetic field sensors, magnetic field-tuneable phase retarders, and wavelength-tuneable optical isolators.

**References**

1  Frisk Kockum, A., Miranowicz, A., De Liberato, S., Savasta, S. & Nori, F. Ultrastrong coupling between light and matter. *Nat. Rev. Phys* **1**, 19-40 (2019).
2  Zheng, Z.-g. *et al.* Three-dimensional control of the helical axis of a chiral nematic liquid crystal by light. *Nature* **531**, 352 (2016).
3  Siegrist, F. *et al.* Light-wave dynamic control of magnetism. *Nature* **571**, 240–244 (2019).
4  Lee, H.-E. *et al.* Amino-acid- and peptide-directed synthesis of chiral plasmonic gold nanoparticles. *Nature* **556**, 360-365 (2018).
5  Goodling, A. E. *et al.* Colouration by total internal reflection and interference at microscale concave interfaces. *Nature* **566**, 523-527 (2019).
6  Horng, Herng-Er, et al. "Magnetochromatics of the magnetic fluid film under a dynamic magnetic field." *Appl. Phys. Lett.* **79**, 350-352 (2001).
7  Walba, D. M. *et al.* A Ferroelectric Liquid Crystal Conglomerate Composed of Racemic Molecules. *Science* **288**, 2181 (2000).
8  Link, D. R. *et al.* Spontaneous Formation of Macroscopic Chiral Domains in a Fluid Smectic Phase of Achiral Molecules. *Science* **278**, 1924 (1997).
9  Zhang, G., Peng, W., Wu, J., Zhao, Q. & Xie, T. Digital coding of mechanical stress in a dynamic covalent shape memory polymer network. *Nat. Commun.* **9**, 4002 (2018).
10  Blachnik, N., Kneppe, H. & Schneider, F. Cotton-Mouton constants and pretransitional phenomena in the isotropic phase of liquid crystals. *Liq. Cryst.* **27**, 1219-1227 (2000).
11  Szczytko, J., Vaupotič, N., Osipov, M. A., Madrak, K. & Górecka, E. Effect of dimerization on the field-induced birefringence in ferrofluids. *Phys. Rev. E* **87**, 062322 (2013).
12  Martinez, L., Cecelja, F. & Rakowski, R. A novel magneto-optic ferrofluid material for sensor applications. *Sens. Actuators, A* **123-124**, 438-443 (2005).
13  Gong, C. *et al.* Discovery of intrinsic ferromagnetism in two-dimensional van der Waals crystals. *Nature* **546**, 265 (2017).

14  Bonilla, M. *et al.* Strong room-temperature ferromagnetism in VSe2 monolayers on van der Waals substrates. *Nat. Nanotechnol.* **13**, 289-293 (2018).
15  Huang, B. *et al.* Layer-dependent ferromagnetism in a van der Waals crystal down to the monolayer limit. *Nature* **546**, 270 (2017).
16  Deng, Y. *et al.* Gate-tuneable room-temperature ferromagnetism in two-dimensional Fe3GeTe2. *Nature* **563**, 94-99 (2018).
17  Ge, Jianping, Yongxing Hu, and Yadong Yin. "Highly tuneable superparamagnetic colloidal photonic crystals." *Angew. Chem. Int. Ed.* **46**, 7428 (2007).
18  Kim, Hyoki, et al. "Structural colour printing using a magnetically tuneable and lithographically fixable photonic crystal." *Nat. Photonics* **3**, 534 (2009).
19  Sano, Koki, et al. "Photonic water dynamically responsive to external stimuli." *Nat. Commun.* **7**, 12559, (2016).
20  Tseng, P., Napier, B., Zhao, S., Mitropoulos, A. N., Applegate, M. B., Marelli, B., & Omenetto, F. G. Directed assembly of bio-inspired hierarchical materials with controlled nanofibrillar architectures. *Nat. Nanotechnol.*, **12**, 474 (2017).
21  Beaufort, Luc, Nicolas Barbarin, and Yves Gally. "Optical measurements to determine the thickness of calcite crystals and the mass of thin carbonate particles such as coccoliths." *Nat. Protoc.* **9**, 633 (2014).
22  Lin, Jing-Fung, and Meng-Zhe Lee. "Concurrent measurement of linear birefringence and dichroism in ferrofluids using rotating-wave-plate Stokes polarimeter." *Opt. Commun.* **285**, 1669-1674 (2012).
23  Patel, Rajesh, V. K. Aswal, and Ramesh V. Upadhyay. "Magneto-optically induced retardation and relaxation study in a mixed system of magnetic fluid and cationic micelles." *J. Magn. Magn. Mater.* **320**, 3366-3369 (2008).
24  Fang, Yimei, et al. "Large magneto-optical effects and magnetic anisotropy energy in two-dimensional Cr$_2$Ge$_2$Te$_6$." *Phys. Rev. B* **98**. 125416 (2018).
25  Burch, K. S., Mandrus, D. & Park, J.-G. Magnetism in two-dimensional van der Waals materials. *Nature* **563**, 47-52 (2018).
26  Shen, T.-Z., Hong, S.-H. & Song, J.-K. Electro-optical switching of graphene oxide liquid crystals with an extremely large Kerr coefficient. *Nat. Mater.* **13**, 394 (2014).
27  Pearson, R. F. "Magnetic anisotropy in ferrimagnetite crystals". *J. Appl. Phys.* **31**, S160 (1960).
28  Kielich, Stanisław. "Frequency doubling of laser light in an isotropic medium with electrically destroyed centre of inversion." *Opto-electronics* **2**, 5-20 (1970).
29  Mertelj, Alenka, et al. "Ferromagnetism in suspensions of magnetic platelets in liquid crystal." *Nature* **504** 237 (2013).